

\magnification=1200

\font\pl=cmssq8 scaled \magstep1

\font\sf=cmss10
\font\titlesf=cmssbx10 scaled \magstep2
\font\chaptsf=cmssbx10
\font\authorsf=cmss10 scaled \magstep1

\def\RR{\rm I\!R}

\def\d{\delta}

\def\l{\lambda}

\def\ltextindent#1{\hbox to \hangindent{#1\hss}\ignorespaces}

\headline={\ifodd\pageno\rightheadline\else\leftheadline\fi}
\def\rightheadline{\it\title\qquad\hfill\rm\folio}
\def\leftheadline{\rm\folio\hfill\it\qquad\author}

%

\nopagenumbers
\def\author{I. G. Avramidi and R. Schimming}
\def\title{Heat kernel coefficients to the matrix Schr\"odinger operator}

{\nopagenumbers
\null
\headline={}
\vskip-1.5cm
\hskip7.5cm{ \hrulefill }
\vskip-.55cm
\hskip7.5cm{ \hrulefill }
\smallskip
\rightline{{\pl University of Greifswald (1994)}}
\rightline{hep-th/9501026}
\smallskip
\hskip7.5cm{ \hrulefill }
\vskip-.55cm
\hskip7.5cm{ \hrulefill }
\bigskip
\hskip7.5cm{submitted to:}

\hskip7.5cm{\sf Journal of Mathematical Physics}

\vfill

\centerline{\titlesf Heat kernel coefficients to}
\medskip
\centerline{\titlesf the matrix Schr\"odinger operator}
\bigskip

\centerline{{\authorsf I. G. Avramidi}
\footnote{$^{a)}$}{ Alexander von Humboldt Fellow}
\footnote{$^{b)}$}{On leave of absence from Research Institute for Physics,
Rostov State University, Stachki 194, Rostov-on-Don 344104, Russia}
\footnote{$^{c)}$}{E-mail: avramidi@math-inf.uni-greifswald.d400.de}
{\authorsf and R. Schimming}
\footnote{$^{d)}$}{E-mail: schimmin@math-inf.uni-greifswald.d400.de}}

\centerline{\it Department of Mathematics, University of Greifswald}
\centerline{\it Jahnstr. 15a, 17489 Greifswald, Germany}
\bigskip\smallskip
\vfill

\bigskip
{\narrower
{
The heat kernel coefficients $H_k$ to the Schr\"odinger operator with
a matrix potential are investigated. We present algorithms and explicit
expressions for the Taylor coefficients of the $H_k$. Special terms are
discussed, and for the one-dimensional case some improved algorithms
are derived.

}}
\eject}


\leftline{\chaptsf I. INTRODUCTION}
\bigskip

Let us consider the Laplace operator on functions over some domain
$M\subseteq \RR^n$ with Cartesian coordinates $x^a=x^1,\dots,x^n$:
$$
\Delta=g^{ab}\partial_a\partial_b, \qquad
\partial_a={\partial\over\partial x^a},
\eqno(1.1)
$$
with a constant positive definite metric $(g_{ab})=(g^{ab})^{-1}$, and
the matrix Schr\"o\-dinger operator
$$
L=I\Delta +U,
\eqno(1.2)
$$
where $I$ is the $N\times N$ unit matrix and $U=U(x)$ some $N\times N$
matrix-valued potential. We assume the potential $U$ to be Hermitian,
$U^{\dag}=U$, and smooth in the domain $M$. If $M$ is noncompact, it
is assumed, in addition, that $U(x)$ decreases sufficiently rapidly at
infinity.

It is well known that $L$ is an elliptic and self-adjoint operator with
a real spectrum bounded from above on a Hilbert space of functions over
$M$ with some suitable boundary conditions. In the case of a periodic
potential the boundary conditions can be replaced by periodicity conditions,
and in the case of a noncompact domain they take the form of some asymptotic
conditions at infinity, requiring sufficiently fast decrease.

This implies that for  any $t>0$ the operator $K(t)=\exp(tL)$ is well defined.
Its kernel $K(x,x';t)$, often referred to as the heat kernel, is the
fundamental solution of  the heat equation
$$
(I\partial_t -L) K=0;
\eqno(1.3)
$$
it satisfies a singular initial condition
$$
K(x,x';+0)=({\rm det}g_{ab})^{-1/2}\d(x-x')I,
\eqno(1.4)
$$
where $\d(x)$ denotes the Dirac delta-distribution.
It is well known that when the points $x$ and $x'$ do not lie on the boundary,
there holds an asymptotic expansion as $t\to +0$ near the diagonal
$x\approx x'$ of the form
$$
K(x,x';t)\sim (4\pi t)^{-n/2}
\exp\left\{-{1\over 4t}\vert x-x'\vert^2\right\}
\sum\limits_{k=0}^\infty t^k H_k(x,x'),
\eqno(1.5)
$$
where $|x|=g_{ab}x^ax^b$.
{}From (1.3) and (1.5) it follows that the sequence of the coefficients $H_k$
is characterized by the differential-recursion relations
$$
\left(D +k \right) H_k = L H_{k-1}, \qquad {\rm for}\
k\ge 1,
\eqno(1.6)
$$
where we abbreviate
$$
D=(x-x')^a\partial_a,
\eqno(1.7)
$$
together with the initial condition
$$
H_0=I.
\eqno(1.8)
$$

The analysis of the heat kernel, or Hadamard, coefficients
$H_k=H_k(x,x'), (k=0,1,2\dots),$ introduced by (1.6)-(1.8) is the
subject of the present paper.
Their role goes far beyond the heat kernel. They can be introduced for any
linear second order differential operator with a Laplace-like principal
part (cf. e.g.$^{1-5}$). The sequence ($H_k$) turns out to have diverse
mathematical and physical applications: spectral geometry, Hadamard's
elementary solution, Huygens' principle, statistical physics,
determinants of elliptic operators, anomalies, Korteweg-de Vries
hierarchy of soliton equations etc.

In the present paper we assume the metric ($g_{ab}$) to be Euclidean,
i.e. positive definite, and the operator $L$ to be elliptic for the sake
of simplicity. The formal results hold for a pseudo-Euclidean metric of
any signature and hyperbolic (normally hyperbolic or ultrahyperbolic)
operators too.

Our paper is organized as follows. In Sects. II and III we obtain
explicit formulas for the Taylor coefficients $[d^pH_k] (p=0,1,2,\dots)$
of $H_k(x,x')$, where $x$ is the variable point and $x'$ is the origin.
We analyse these quantities and determine some special terms and a `common
denominator' (see below). In Sect. IV we consider the one-dimensional case and
derive a simplified algorithm for the diagonal values $[H_k]$ of the $H_k$ via
a matrix Riccati equation. Then we generalize the so-called `Lenard recursion'
from the scalar case to the matrix one. The new recursion is formally
non-local, but can be effectively done in the space of differential polynomials
in $U$. It leads to a proposal of some matrix Korteweg-de Vries hierarchy for
$U=U(x,\tau)$.

It is not possible to quote all the literature on the heat kernel approach. Let
us mention papers which deal with the matrix case$^{6-12}$. The methods applied
there differ from ours.

\bigskip
\bigskip
\leftline{\chaptsf II. THE HEAT KERNEL COEFFICIENTS}
\vglue0pt
\bigskip
\vglue0pt
For a smooth potential $U=U(x)$ the coefficients $H_k(x,x')$ are known to be
smooth near the diagonal $x=x'$. Thus, to study the sequence $(H_k)$ we use the
Taylor expansion with respect to the variable $x$ at the origin $x'$ taken as
an asymptotic series as $x\to x'$:
$$
H_k(x,x')\simeq\sum\limits_{p=0}^\infty {1\over p!}(x-x')^{a_1}\cdots
(x-x')^{a_p}
[\partial_{a_1}\cdots\partial_{a_p}H_k].
\eqno(2.1)
$$
Here and in the following the square brackets $[ \ ]$ denote the diagonal value
of a two-point quantity $H=H(x,x')$:
$$
[H]=[H](x):=H(x,x).
\eqno(2.2)
$$

The coefficients of the Taylor expansion can be arranged to matrix-valued
symmetric differential forms (or covariant tensors), shortly $p$-forms,.
The symmetric product $\vee$ of two such forms $A_{(p)}$ and $B_{(q)}$ is
defined as
$$
A_{(p)}\vee B_{(q)}=A_{(a_1\cdots a_p}B_{b_1\cdots b_q)} dx^{a_1}\otimes\cdots
\otimes dx^{a_p}\otimes dx^{b_1}\otimes\cdots \otimes dx^{b_q},
\eqno(2.3)
$$
where the parentheses mean symmetrization over all indices included. The
symmetric differential
$$
d: A_{(p)}\to A_{(p+1)}
\eqno(2.4)
$$
on these forms is defined by
$$
dA_{(p)}=\partial_{(a_1}A_{a_2\cdots a_{p+1})}
dx^{a_1}\otimes\cdots \otimes dx^{a_{p+1}}.
\eqno(2.5)
$$
Using these notations, we have
$$
[d^pH_k]= [\partial_{a_1}\cdots\partial_{a_p} H_k] dx^{a_1}\otimes\cdots\otimes
dx^{a_p}.
\eqno(2.6)
$$

Further, we define some trace operation
$$
{\sf g^{-1}}: A_{(p)}\to A_{(p-2)}
\eqno(2.7)
$$
by
$$
\eqalignno{
g^{-1}A_{(0)}&=0, \qquad g^{-1}A_{(1)}=0,&\cr
{\sf g^{-1}}A_{(p)}&=g^{a_1a_2}A_{(a_1a_2a_3\cdots a_p)}dx^{a_3}\otimes\cdots
\otimes dx^{a_{p}},\qquad {\rm for\ }p\ge 2. &(2.8)\cr}
$$
The Laplacian on symmetric forms
$$
\Delta:={\sf g^{-1}}d^2: \ A_{(p)}\to A_{(p)}
\eqno(2.9)
$$
generalizes the usual Laplacian (1.1) on functions, i.e. $0-$forms.

Let us now apply $d^p$ to (1.6) and then restrict to the diagonal $x=x'$.
Taking into account the commutation relation
$$
[d^p, D]=pd^p,
\eqno(2.10)
$$
we obtain the purely algebraic recursion system
$$
(p+k)[d^pH_k]={\sf g^{-1}}[d^{p+2}H_{k-1}]+
\sum\limits_{q=0}^{p}{p\choose q} U_{(p-q)}
\vee [d^{q}H_{k-1}], \qquad {\rm for \ }
k\ge 1,\qquad p\ge 0
\eqno(2.11)
$$
where $U_{(p)}=d^{p}U$,
with the initial conditions
$$
\eqalignno{
[H_0]&=I, \qquad [d^pH_0]=0 \qquad {\rm for}\ p\ge 1.&(2.12)\cr}
$$

An explicit non-recursive solution of (2.11), (2.12) can be presented by means
of the following symbolism. Let us define linear operators $M_q,
(q=-2,-1,0,1,\dots )$ on matrix-valued symmetric forms $A_{(p)}$ according to
$$
\eqalignno{
M_{-2}A_{(p)}&:={\sf g^{-1}}A_{(p)}, \qquad M_{-1}A_{(p)}:=0,&(2.13)\cr
M_q A_{(p)}&:={p+q\choose q}U_{(q)} \vee A_{(p)},\qquad {\rm for\ } q\ge 0.
&(2.14)\cr}
$$
Then the system (2.11) can be rewritten in a short-hand notation:
$$
[d^pH_k]=\sum\limits_{q=0}^{p+2} {1\over p+k}M_{p-q}[d^qH_{k-1}] .
\eqno(2.15)
$$
By applying mathematical induction to this recursion system we obtain
the following.

\smallskip
\noindent
{\bf Theorem 2.1.}
{\sl There holds for $k\ge 1, p\ge 0$
$$
[d^pH_k]=\sum\limits_{q_2,\dots,q_k}
\left(\prod\limits_{r=1}^k{1\over (q_r+k-r+1)}M_{q_r-q_{r+1}}\right)I,
\eqno(2.16)
$$
where the sum runs over the integers $q_2, q_3,\dots,q_k$ so that
$$
0\le q_{r+1}\le q_r+2, \qquad {\rm for\ } r=1,2,\dots, k-1,
$$
and
$$
q_1=p,\qquad q_{k+1}=0.
$$
}
\smallskip

This result can also be obtained, as a special case, from the general explicit
formula for the heat kernel coefficients for any second-order elliptic operator
of Laplace type obtained by one of the authors (I.G.A.) in$^5$.

The formula (2.16) is not well suited to recognize special polynomial
constituents of $[d^pH_k]$. To do this, we decompose $[d^pH_k]$ into its
homogeneous parts of {\it degree} $d$ with respect to $U$:
$$
[d^pH_k]=\sum\limits_{d=1}^{k}H_{k, d}^{(p)}.
\eqno(2.17)
$$

The system (2.11) decomposes in an easy way into homogeneous parts too. Thereby
the linear and the highest degree parts decouple from the whole system:
$$
\eqalignno{
(p+k)H_{k,1}^{(p)}&={\sf g}^{-1}H_{k-1, 1}^{(p+2)}, &(2.18)\cr
(p+k)H_{k,k}^{(p)}&=\sum\limits_{q=0}^{p}{p\choose q}U_{(p-q)}
\vee H_{k-1,k-1}^{(q)}. &(2.19) \cr}
$$
The initial values are given by
$$
(p+1)H_{1,1}^{(p)}=U_{(p)}.
\eqno(2.20)
$$
Mathematical induction gives the following.
\smallskip
\noindent
{\bf Theorem 2.2.}
{\sl The linear part $H_{k,1}^{(p)}$ of $[d^pH_k]$ equals
$$
H_{k,1}^{(p)}={ (p+k-1)!\over (p+2k-1)!} d^p\Delta^{k-1} U,
\eqno(2.21)
$$
while the highest degree part of $H_{k,k}^{(p)}$ of $[d^pH_k]$ is
$$
H_{k,k}^{(p)}=\sum\limits_{q_2,\cdots,q_k}
\bigvee\limits_{r=1}^{k}
{1\over q_r+k-r+1}{q_r\choose q_{r+1}}U_{(q_r-q_{r+1})},
\eqno(2.22)
$$
where the sum runs over the integers $q_2, q_3,\cdots q_k$, so that
$$
0\le q_{r+1}\le q_r \qquad {\rm for\ }\ r=1,2,\dots, k-1
$$
$$
q_1=p,\qquad q_{k+1}=0.
$$
}
Let us call a number-theoretical function $d(p,k)$ a {\it common denominator}
to the system of differential polynomials $[d^pH_k]$ if all the modified
polynomials in $U, U_{(1)}, U_{(2)}, \dots $
$$
\bar H^{(p)}_k:=d(p,k)[d^pH_k]
\eqno(2.23)
$$
have integer coefficients.

\smallskip
\noindent
{\bf Proposition 2.3.}
{\sl A common denominator in the aforesaid sense is given by
$$
d(p,k)={1\over p!}\prod\limits_{j=0}^{l-1}{(p+2k-2j-1)!\over (2j)!},\qquad {\rm
for\ even\ } k=2l,
\eqno(2.24)
$$
and
$$
d(p,k)={p+k\over p!}\prod\limits_{j=0}^{l-1}{(p+2k-2j-1)!\over (2j)!},\qquad
{\rm for\ odd\ } k=2l-1,
\eqno(2.25)
$$
}
\noindent
{\bf Proof.}
Substituting
$$
[d^pH_k]={1\over d(p,k)}\bar H^{(p)}_k
\eqno(2.26)
$$
into (2.11) we get the recursion system for the $\bar H_k^{(p)}$
$$
\bar H^{(p)}_k={d(p,k)\over (p+k)d(p+2,k-1)}
{\sf g^{-1}}\bar H^{(p+2)}_{k-1}+
\sum\limits_{q=0}^{p}{p \choose q}{d(p,k)\over (p+k)d(q,k-1)} U_{(p-q)}
\vee \bar H^{(q)}_{k-1},
\eqno(2.27)
$$
for $k\ge 1, p\ge 0$.
One can show that the coefficients of this system are integer, which proves the
theorem.
\smallskip

To find the {\it least} common denominator remains an open problem.
Fulling$^{13}$ found the expressions (2.24), (2.25) except for the factor
${1/p!}$. It is obvious, that this factor does not matter for $p=0$.

\bigskip
\bigskip
\leftline{\chaptsf III. THE DIFFERENTIAL POLYNOMIALS $[H_k]$}
\vglue0pt
\bigskip
\vglue0pt
Let us analyse more closely the structure of the differential polynomials
$[H_k]$.
The common denominator for the system $[H_k], (k=1, 2, \cdots)$ reads
$$
d(0,k)=\prod_{j=0}^{[(k-1)/2]}{(2k-2j-1)!\over (2j)!}.
\eqno(3.1)
$$
The extremal terms are given by
$$
[H_k]={(k-1)!\over (2k-1)!} \Delta^{k-1}U+\cdots
+{1\over k!}U^k, \qquad {\rm for\ }k\ge 1,
\eqno(3.2)
$$
where the points indicate terms of a degree higher than $1$ and less than $k$.
The sequence begins with
$$
\eqalignno{
[H_1]&=U,&\cr
[H_2]&={1\over 6}\left(\Delta U+3U^2\right),&(3.3)\cr
[H_3]&={1\over 60}\left\{\Delta^2U+5\left[U(\Delta U)
+(\Delta U) U + {\sf g}^{-1} (dU\vee dU)\right]+10 U^3
\right\}.&\cr}
$$
Using the identity
$$
2{\sf g}^{-1}(dW_1\vee dW_2)=\Delta (W_1 W_2)-(\Delta W_1)W_2
-W_1(\Delta W_2),
\eqno(3.4)
$$
whith $W_1, W_2$ being $0$-forms , one can eliminate the operator ${\sf
g}^{-1}$ here:
$$
[H_3]={1\over 120}\left\{2\Delta^2U + 5U(\Delta U)
+5(\Delta U) U + 5\Delta (U^2) + 20 U^3
\right\}.
\eqno(3.5)
$$

Let us call a monomial or differential polynomial {\it simple} if it is solely
composed from $U$ and $\Delta$ (but not from $d$ and $g^{-1}$), and the total
number of symbols $U$ and $\Delta$ in a simple monomial the {\it weight} $w$.
It is obvious that a simple monomial is always a $0$-form and that the number
of symbols $U$ in it equals its degree.

\smallskip\noindent
{\bf Theorem 3.1.}
{\sl There are exactly
$$
C(w,d)={1\over w}{w \choose d}{w \choose d-1}
\eqno(3.6)
$$
simple monomials of degree $d$ and weight $w$.}

\noindent
{\bf Proof.}
A simple monomial $W$ reads
$W=U^w$ or
$$
W=U^{d_0}(\Delta W_1)W_2,
\eqno(3.7)
$$
where $U^{d_0}$ and $W_2$ may be empty, while $W_1$ is not.
Since the decomposition is unique, the number $C(w,d)$ of monomials $W$ follows
from the numbers $C(w_1,d_1), C(w_2,d_2)$ of $W_1, W_2$ respectively. Thus,
$$
C(w,d)=\sum\limits_{d_0=0}^{w-1}\sum\limits_{d_1,d_2;w_1,w_2}
C(w_1,d_1)C(w_2,d_2)
\eqno(3.8)
$$
for $d<w$, where the inner sum runs over integers $w_1,w_2,d_1,d_2$ so that
$$
w_1\ge d_1\ge 1,\qquad w_2\ge d_2\ge 0,\qquad w_1+w_2=w-d_0-1,\qquad
d_1+d_2=d-d_0.
$$
The recursion formula (3.8) can be translated into a quadratic equation
$$
xyF^2+(x+y-1)F+1=0
\eqno(3.9)
$$
for the generating function
$$
F=F(x,y)=\sum\limits_{w=1}^{\infty}\sum\limits_{d=1}^{w}C(w,d)x^{d-1}y^{w-d},
\eqno(3.10)
$$
with the initial condition
$$
F(0,0)=1.
\eqno(3.11)
$$
Further, using
$$
(1+\sqrt{1-4z})^{-1}={1\over 4}\sum\limits_{w=0}^\infty {1\over w+1}{2w\choose
w}z^w
\eqno(3.12)
$$
we expand the solution of the equation (3.9)
$$
F(x,y)=2\left(1-x-y+\sqrt{(1-x-y)^2-4xy}\right)^{-1}
\eqno(3.13)
$$
in the power series (3.10) and prove that (3.6) is correct.
\smallskip
B. Fiedler$^{14}$ derived special cases of theorem 3.1, namely for $w=2$ and
$w=3$, by combinatorial arguments.

Now, let us try to eliminate ${\sf g}^{-1}$ from the differential polynomials
by means of the identities like (3.4). Making use of
$$
2{\sf g}^{-1}(dW_1 \vee W_2 \vee dW_3)=\Delta (W_1W_2W_3)-(\Delta (W_1W_2))W_3
- W_1\Delta (W_2W_3)+W_1(\Delta W_2)W_3.
\eqno(3.14)
$$
where $W_i$ are $0$-forms, we easily get a simple form for $[H_4]$:
$$
\eqalignno{
[H_4]&={1\over 120}\Biggl\{{1\over 7}\Delta^3 U+ {1\over 3}\left[\Delta^2 U^2
+ (\Delta^2 U)U+U\Delta^2U+(\Delta U)^2+\Delta((\Delta U)U)
+\Delta(U(\Delta U))
\right]&\cr
&+\Delta (U^3)+(\Delta (U^2))U+U\Delta (U^2)+(\Delta U)U^2+U^2\Delta U+
U(\Delta U)U+5U^4
\Biggr\}. &(3.15)\cr}
$$
However, it is impossible to do this for higher polynomials $[H_k], k\ge 5,$
because of terms like
$$
g^{ac}g^{bd}(\partial_a W_1)(\partial_b W_2)(\partial_c W_3)(\partial_d W_4).
$$

One of the authors (R.Sch.) has calculated $[H_1], \dots, [H_4]$ in simple form
for the matrix Schr\"odinger operator in the early paper$^{15}$. Later T.
Osborn et al.$^8$ obtained the result in a "non-simple" form. The other author
(I.G.A.) calculated the coefficients $[H_1], \dots, [H_4]$ in the general case,
that means for every Laplace-type operator, in$^{16,17,5}$.

\bigskip
\bigskip
\leftline{\chaptsf IY. THE ONE-DIMENSIONAL CASE}
\bigskip

Let us discuss more detailed the one-dimensional Schr\"odinger operator
$$
L=d^2+U,
\eqno(4.1)
$$
where now $d=d/dx$ and n=1.
In this case the recursion system (2.11), (2.12) reduces to
$$
(p+k)[d^pH_k]=[d^{p+2}H_{k-1}]+\sum\limits_{q=0}^{p}{p\choose q}U_{p-q}
[d^qH_{k-q}] \qquad {\rm for} \ k\ge 1, p\ge 0
\eqno(4.2)
$$
$$
[H_0]=I, \qquad [d^pH_0]=0, \qquad {\rm for \ } p\ge 1.
$$
The explicit solution (2.16 ) of this system can be rewritten now in a better
form.

\smallskip
\noindent
{\bf Theorem 4.1.}
{\sl There holds for $k\ge 2, p\ge 0$
$$
[d^pH_k]=\sum\limits_{q_2,\cdots ,q_k}\prod\limits_{r=1}^{k}
{c(q_r,q_{r+1})\over q_r+k-r+1}U_{q_r-q_{r+1}},
\eqno(4.3)
$$
where
$$
\eqalignno{
U_{-2}&:=I,\qquad U_{-1}:=0,\qquad U_p:=U_p=d^pU \qquad {\rm for \ } p\ge 0,
&(4.4)\cr
c(p,q)&:={p\choose q}+\d^{p+2}_q,&(4.5)\cr}
$$
and the sum runs over the integers $q_2, q_3,\cdots,q_k$ so that
$$
0\le q_{r+1}\le q_r+2 \qquad {\rm for\ } r=1,2,\cdots,k-1,
$$
$$
q_1=p,\qquad q_{k+1}=0.
$$
}
A proof has been given by R.Sch. in$^{18}$ for the scalar case ($N=1$).
Inspection shows that the arguments of$^{18}$ do not depend on the number $N$
of components and are valid for the matrix case too. Fulling$^{13}$ found that
formula (4.3) is well suited for calculating the quantities $[d^pH_k]$ by means
of computer algebra. This theorem also follows, as a rather special case, from
the results obtained by I.G.A. in$^{5}$, where a general explicit formula for
the heat kernel coefficients to any second-order elliptic operator of Laplace
type is obtained.

 {}From Theorem 4.1. it follows that the quantity $[d^pH_k]$ for $k\ge 1, p\ge
0$ can be brought into the form
$$
[d^pH_k]=\sum\limits_{d=1}^k\sum\limits_{r_1,\dots,r_d\ge
0}a(r_1,r_2,\dots,r_d)
U_{r_1}U_{r_2}\cdots U_{r_d},
\eqno(4.6)
$$
where $a(r_1,r_2,\dots,r_d)$ are some positive coefficients and the inner sum
runs over the integers $r_1,r_2\dots,r_d$ so that
$$
r_1+r_2+\cdots+r_d=p+2(k-d).
$$

There is a one-to-one correspondence between the ordered partitions $(r_1+1,
r_2+1, \cdots ,r_d+1)$ of $s=\sum_{i=1}^d(r_i+1)=p+2k-d$ into $d$ numbers and
the monomials $U_{r_1}U_{r_2}\cdots U_{r_d}$. Since there are exactly
${s-1\choose d-1}$ such partitions, the total number of monomials, i.e. the
total number of the terms in the sum (4.6) equals
$$
\sum\limits_{d=1}^{k}{p+2k-d-1\choose d-1}.
\eqno(4.7)
$$

The formula (4.3) is explicit, i.e. non-recursive, but somewhat complicated
because each coefficient $c(q_r,q_{r+1})$ comprises the two cases $q_{r+1}\le
q_r$ and $q_{r+1}=q_r+2$. As an alternative, we offer a recursion for the
sequence $[H_k], (k=0,1,2,\dots)$ which is by far more effective than the
double recursion (4.2) for the double sequence $[d^pH_k], (p,k=0,1,2,\dots)$.

\smallskip
\noindent
{\bf Theorem 4.2.}
{\sl
Let the differential polynomials $Z_k, (k=1,2,\dots)$ in $U=U(x)$ be
recursively defined by
$$
Z_{k+1}=dZ_k+\sum\limits_{m=1}^{k-1}Z_mZ_{k-m}
\eqno(4.8)
$$
$$
Z_1=U.
\eqno(4.9)
$$
Then the quantities
$$
G_k:={(2k)!\over 2k!}[H_k]
\eqno(4.10)
$$
are determined by the recursion
$$
G_k=2\sum\limits_{m=1}^{k}Z_{2m-1}G_{k-m}, \qquad {\rm for \ } k\ge 1,
\eqno(4.11)
$$
together with the initial condition
$$
G_0={1\over 2}I.
\eqno(4.12)
$$
}
\noindent
{\bf Proof.}
Let us start with the heat kernel $K(x,x';t)$ to $L$. Its Laplace transform
$$
G(x,x';\l)=\int\limits_0^\infty dt\exp(-t\l)K(x,x';t).
\eqno(4.13)
$$
is the resolvent kernel to $L$, i.e. the fundamental solution of $L-\l I$,
which means
$$
(d^2+U(x)-\l I)G(x,x';\l)=-\d(x-x')I.
\eqno(4.14)
$$

The resolvent kernel is continuous but not differentiable at $x=x'$. Near the
diagonal, $x\approx x'$, it has the form
$$
G(x,x';\l)=[G(\l)](x')-{1\over 2}|x-x'|
+(x-x')F(x';\l)+O((x-x')^2),
\eqno(4.15)
$$
where $[G(\l)](x):=G(x,x;\l)$, so that
$$
d G(x,x';\l)= -{1\over 2}{\rm sign}(x-x')+F(x';\l)+O(x-x'),
\eqno(4.16)
$$
where $|x|$ is the absolute value of $x$, ${\rm sign}(x)$ denotes the sign of
$x$ and $F(x';\l)$ is some matrix that does not depend on $x$.
For sufficiently large ${\rm Re} \l$ and $x\approx x'$ the resolvent kernel is
known to be invertible, so that one can present the derivative of the resolvent
in the form
$$
d G(x,x';\l):=Z(x,x';\l)G(x,x';\l).
\eqno(4.17)
$$

Then from (4.16), (4.17) we conclude that $Z=Z(x,x';\l)$ is discontinuous with
a jump at $x=x'$ that determines the resolvent kernel on the diagonal:
$$
\{Z(x'+0,x';\l)-Z(x'-0,x';\l)\}[G(\l)](x')=-I.
\eqno(4.18)
$$

Moreover, the  equation (4.14) for $x\ne x'$ becomes equivalent to
the Riccati equation for $Z$
$$
d Z(x,x';\l)+ Z^2(x,x';\l)+U(x)-\l I=0.
\eqno(4.19)
$$

Let us look for a solution of this equation in the form of an asymptotic
expansion as ${\rm Re\, }\l\to\infty, \vert{\rm arg} \l\vert < \pi$.
Using the asymptotic expansion of the heat kernel (1.5) and (4.13) we obtain
$$
Z(x,x';\l)\sim X(x;\l)+{\rm sign}(x-x')Y(x;\l),
\eqno(4.20)
$$
where the formal power series
$$
\eqalignno{
X(x;\l)&=\sum\limits_{k=0}^\infty (4\l)^{-k}Z_{2k}(x), &(4.21)\cr
Y(x;\l)&=\sum\limits_{k=0}^\infty (4\l)^{-k+1/2}Z_{2k-1}(x),&(4.22)\cr}
$$
are composed of the quantities $Z_k(x)$ defined above.
Namely, substituting the expansion into the Riccati equation (4.19), we find
that $Z_{-1}=I, Z_0=0$ and $Z_k(x), (k=1,2,\dots),$ are just the differential
polynomials in $U(x)$ determined by the recursion system (4.8), (4.9).

Therefore, the asymptotic expansion of $Z(x,x';\l)$ is purely local but
discontinuous at $x=x'$. From (4.20) it is clear that the jump of $Z(x,x';\l)$
at $x=x'$ equals just $Y(x;\l)$, so that (4.18) takes the form
$$
2Y(x;\l)[G(\l)](x)=I
\eqno(4.23)
$$
Using (4.13) and the asymptotic expansion of the heat kernel on the diagonal as
$t\to +0$
$$
[K(t)]\sim(4\pi)^{-1/2}\sum\limits_{k=0}^\infty t^{k-1/2}[H_k],
\eqno(4.24)
$$
we easily get the corresponding asymptotic expansion of the resolvent on the
diagonal as $\l\to\infty$:
$$
[G(\l)]\sim 2\sum\limits^\infty_{k=0}(4\l)^{-k-1/2} G_k,
\eqno(4.25)
$$
where
$$
G_k:={(2k)!\over 2k!}[H_k].
\eqno(4.26)
$$
Therefrom, using (4.22) and (4.23), we find that the coefficients $G_k$ are
determined by the recursion system (4.11), which proves the theorem.

The result of the Theorem 4.2. can be also obtained from the results of the
paper of Fulling$^7$, who uses a different method and notation.

\smallskip
 One can explicitly find the linear and quadratic parts of the differential
polynomials $Z_k$ and $G_k$.

\noindent
{\bf Theorem 4.3.}
{\sl There holds for $k\ge 3$
$$
Z_k=U_{k-1}+\sum\limits^{k-3}_{p=0}\left\{{k-1\choose
p+1}-1\right\}U_pU_{k-3-p}+\cdots,
\eqno(4.27)
$$
where the points indicate terms of degrees higher than two.
}

\noindent
{\bf Proof.}
The linear $Z_{k,1}$ and quadratic $Z_{k,2}$ parts of $Z_k$ satisfy the
recursion systems
$$
\eqalignno{
Z_{k+1,1}&=dZ_{k,1},\qquad
Z_{1,1}=U,&(4.28)\cr
Z_{k+1,2}&=dZ_{k,2}+\sum\limits^{k-1}_{m=1}Z_{m,1}Z_{k-m,1},\qquad
Z_{3,2}=U_2.&(4.29)\cr}
$$
Mathematical induction gives (4.27).

\smallskip
\noindent
{\bf Conclusion 4.4.}
{\sl There holds for $k\ge 2$
$$
G_k=U_{2k-2}+\sum\limits^{2k-4}_{p=0}\left\{{2k-2\choose
p+1}+(-1)^p\right\}U_pU_{2k-4-p}+\cdots + {1\over 2}{2k\choose k}U^k,
\eqno(4.30)
$$
where the points indicate terms of degrees higher than two and less than $k$.
}

\noindent
{\bf Proof.}
The linear $G_{k,1}$ and the highest degree $G_{k,k}$ parts of $G_k$ follow
immediately from (4.11). Therefrom we have also for the quadratic part
$G_{k,2}$
$$
G_{k,2}=2\sum\limits_{m=1}^{k-1}Z_{2m-1,1}G_{k-m,1}+Z_{2k-1,2}.
\eqno(4.31)
$$
Explicit calculation gives the result.

\smallskip

Note that in the scalar case $G_{k,2}$ had been found by means of a completely
different method by Gilkey$^{19}$.


Let us  now derive another recursion system for the differential polynomials
$G_k$.
Let in the following the square and the curly brackets denote the commutator
and anticommutator of matrices respectively.
\smallskip
\noindent
{\bf Theorem 4.5.}
{\sl There exist unique differential polynomials
$$
W_k=:d^{-1}[G_k,U],
\eqno(4.32)
$$
to the sequence $G_k (k=1,2,\dots)$ of differential polynomials in $U$ defined
by (4.11), (4.8), (4.9), so that
$$
[G_k,U]=dW_k,
\eqno(4.33)
$$
and the $G_k$ satisfy the differential-recursion system
$$
dG_{k+1}=d^3G_k+\{Ud+dU, G_k\}+[d^{-1}[G_k,U],U].
\eqno(4.34)
$$
}
\noindent
{\bf Proof.}
Let us consider once more the equation for the resolvent kernel
$$
d^2G(x,x';\l)+(U(x)-\l I)G(x,x';\l)=-\d(x-x')I.
\eqno(4.35)
$$
As assumed from the beginning, the matrix-valued potential is Hermitian,
$U^{\dag}=U$,
so that the Schr\"odinger operator is self-adjoint, $L^{\dag}=L$, and the
resolvent satisfies the symmetry condition
$$
G^{\dag}(x,x';\l)=G(x',x;\l).
\eqno(4.36)
$$
Using this property and interchanging the variables $x$ and $ x'$ we get
another differential equation for $G$:
$$
d'^2G(x,x';\l)+G(x,x';\l)(U(x)-\l I)=-\d(x-x')I.
\eqno(4.37)
$$
where $d'=d/dx'$.

The difference of (4.35) and (4.37) reads
$$
(d+d')(d-d')G(x,x';\l)=G(x,x';\l)U(x')-U(x)G(x,x';\l).
\eqno(4.38)
$$
For every smooth two-point function $H(x,x')$ there holds
$$
[(d+d')H]=d[H].
\eqno(4.39)
$$
Using this rule and restricting (4.38) to the diagonal $x=x'$, we get
$$
[[G],U]=d[(d-d')G],
\eqno(4.40)
$$
or put in another way
$$
d^{-1}[[G],U]=[(d-d')G].
\eqno(4.41)
$$
Both $[G(\l)]$ and $W(\l):=[(d-d')G(\l)]$ admit asymptotic expansions in powers
of $\l^{-1/2}$, where the coefficients $G_k$ and $W_k$ respectively are
differential polynomials in $U$. Hence there are unique differential
polynomials $W_k$ such that
$$
dW_k=[G_k,U].
\eqno(4.42)
$$

Now, let us apply the operators $(d+3d')$ and $(d'+3d)$ to the equations (4.35)
and (4.37) respectively and then take the sum. Restricting to the diagonal the
result
$$
\eqalignno{
&(d+d')^3G(x,x';\l)-4\l(d+d')G(x,x';\l)&\cr
&+(dU(x))G(x,x';\l)+G(x,x';\l)d'U(x')&\cr
&+2U(x)(d+d')G(x,x';\l)+2((d+d')G(x,x';\l))U(x')
&\cr&
-U(x)(d-d')G(x,x';\l)+((d-d')G(x,x';\l))U(x')=0, &(4.43)\cr}
$$
and using (4.39), we obtain
$$
d^3[G]-4\l d[G]+2\{U,d[G]\}+\{dU,[G]\}
+[d^{-1}[[G],U],U]=0
\eqno(4.44)
$$
Substitution of the asymptotic expansion of the resolvent kernel on the
diagonal (4.25) proves that the quantities $G_k$ satisfy the recursion system
(4.34).

\smallskip

The formula (4.34) has been found first by Olmedilla et al.$^{10}$, (see also
the recent paper of Bilal$^{11}$).
In the scalar case, when all commutators vanish, it is known under the name
`Lenard recursion'. As a matter of fact, Burchnall and Chaundy$^{20}$
discovered the sequence $(G_k)$ and the scalar `Lenard recursion' already in
the twenties.

Let us write (4.34) in the form
$$
dG_{k+1}=BG_k
\eqno(4.45)
$$
where the linear operator $B$ is defined by
$$
BH:=d^3H+2\{U,dH\}+\{dU,H\}+[d^{-1}[H,U],U]
\eqno(4.46)
$$
Since the operator $d$ is formally anti-selfadjoint, $d^{\dag}=-d$, so is $B$:
$$
B^{\dag}=-B.
\eqno(4.47)
$$
In the scalar case all the commutators vanish and $B$ becomes a differential
operator
$$
B=d^3+2Ud+2dU,
\eqno(4.48)
$$
where the factor $U$ means multiplication by $U$.
The recursion system (4.45) can be  formally solved:
$$
G_k={1\over 2}(d^{-1}B)^k \cdot I.
\eqno(4.49)
$$
Thereby, the action of the positive powers of the operator $B$ on the identity
matrix is well-defined and produces the differential polynomials $G_k$.


A special motivation for the study of the one-dimensional Schr\"odinger
equation is its relation to the Korteweg-de Vries (KdV) hierarchy. If one
assumes a one-parameter dependence of the potential  $U=U(x,\tau)$, then the
differential polynomials $G_k=G_k[U] (k=1,2,\dots)$ constructed above
constitute the KdV hierarchy in the scalar case:
$$
{\partial\over\partial \tau} U ={\partial\over\partial x} G_k[U].
\eqno(4.50)
$$
For $k=1$ a linear differential equation emerges, for $k=2$ we have the KdV
equation itself, for $k\ge 2$ the equation (4.50) is called a higher KdV
equation. Moreover, it is known that (in the scalar case again) the quantities
$G_k$ as well as the $Z_k$ are conserved densities to the KdV equation.
Several papers discuss a matrix KdV equation$^{12,21,22}$
$$
{\partial\over\partial \tau} U ={\partial\over\partial x}G_2
=(\partial_x)^3U+3(\partial_xU)U+3U(\partial_xU).
\eqno(4.51)
$$

In particular, Olmedilla$^{12}$ proved that (4.51) is solitonic in the sense
that the inverse scattering method works for (4.51). One might guess a
solitonic character for the whole matrix KdV hierarchy (4.50) $(k=2,3,\dots)$.

\bigskip
\bigskip
\leftline{\chaptsf ACKNOWLEDGEMENTS}
\bigskip

The authors are greatly indebted to Bernd Fiedler who supported the work by
computer algebra calculations. Also, we express our gratitude to Stephen
Fulling for valuable hints and discussions. The research of I.G.A. is supported
by the Alexander von Humboldt Foundation. He would also like to thank the
University of Greifswald for its hospitality.

\bigskip
\bigskip
\bigskip

\item{$^1$} J. Hadamard, {\it Lectures on Cauchy's problem in linear partial
	differential equations}, (Yale University Press, New Haven, 1923)

\item{$^2$} P. B. Gilkey,
	{\it Invariance theory, the heat equation and the  Atiyah - Singer
	index theorem} (Publish or Perish, Wilmington, 1984)
\item{$^3$} P. G\"unther, {\it Huygens' Principle and Hyperbolic Equations},
	(Academic Press, San Diego, 1988)
\item{$^4$} R. Schimming, Math. Nachr. {\bf 148}, 145 (1990)
\item{$^5$} I. G. Avramidi, Nucl. Phys. B {\bf 355}, 712 (1991)

\item{$^6$} S. A. Fulling, J. Math. Phys. {\bf 20}, 1202 (1979)
\item{$^7$} S. A. Fulling, SIAM J. Math. Anal. {\bf 14}, 780 (1983)
\item{$^8$} T. A. Osborn, R. A. Corns and Y. Fujiwara, J. Math. Phys. {\bf 26},
453 (1985)
\item{$^9$} F. H. Molzahn and T. A. Osborn, {\it A phase space fluctuation
method for quantum dynamics}, University of Manitoba (1993)
\item{$^{10}$} E. Olmedilla, L. Martinez Alonso and F. Guil,
Nuovo Cim. {\bf 61 B}, 49  (1981)
\item{$^{11}$} A. Bilal, {\it Multi-component KdV hierarchy, $V$-algebra and
	non-Abelian Toda theory}, Princeton University, PUPT-1446, (1994),
	hep-th/9401167
\item{$^{12}$} E. Olmedilla, Inverse Problems {\bf 1}, 219 (1985)
\item{$^{13}$} S. Fulling, personal communication
\item{$^{14}$} B. Fiedler, personal communication
\item{$^{15}$} R. Schimming, Ukrainsk. Mat. \v{Z}urnal {\bf 29}, 351 (1977)
\item{$^{16}$} I. G. Avramidi, Teor. Mat. Fiz., {\bf 79}, 219 (1989)
\item{$^{17}$} I. G. Avramidi, Phys. Lett. {\bf B 238}, 92 (1990)
\item{$^{18}$} R. Schimming, Z. f\"ur Analysis u. ihre Anw. {\bf 7}, 203
(1988)
\item{$^{19}$} P. B. Gilkey, Compos. Math. {\bf 38}, 201 (1979)
\item{$^{20}$} J. L. Burchnall and T. W. Chaundy, Proc. London Math. Soc. (2)
{\bf 21}, 420 (1922)
\item{$^{21}$} F. Calogero and A. Degasperis, Nuovo Cim. {\bf 39 B}, 1 (1977)
\item{$^{22}$} M. Wadati and T. Kamijo, Progr. Theor. Phys. {\bf 52}, 397
(1974)

\bye